\documentclass[3p,preprint,review]{elsarticle}

\usepackage{natbib}
\usepackage{amsfonts}
\usepackage{amsmath}
\usepackage{graphicx}
\usepackage{dcolumn}
\usepackage{soul}
\usepackage{bm}
\usepackage{overpic}
\usepackage{booktabs}
\usepackage{braket}
\usepackage{color}
\usepackage{multirow}
\usepackage[colorlinks,linkcolor=red]{hyperref}
\usepackage{caption}
\begin{document}

\begin{frontmatter}

\title{Accumulative time-based ranking method to reputation evaluation in information networks }

\author[label1,label2]{Hao Liao}
\author[label1]{Qi-xin Liu}
\author[label1]{Ze-cheng Huang}
\author[label3]{Chi Ho Yeung}
\author[label4]{Yi-Cheng Zhang}

\address[label1]{National Engineering Laboratory on Big data System Computing Technology, Guangdong Province Engineering Center of China-made High Performance Data Computing System, College of Computer Science and Software Engineering, Shenzhen University, Shenzhen 518060, PR China}
\address[label2]{Institute of Big Data Intelligent Management and Decision, Shenzhen University, Shenzhen 518060, PR China}
\address[label3]{Department of Science and Environmental Studies, The Education University of Hong Kong, Hong Kong, PR China}
\address[label4]{Department of Physics, University of Fribourg, Fribourg, Switzerland}

\begin{abstract}
With the rapid development of modern technology, the Web has become an important platform for users to make friends and acquire information. However, since information on the Web is over-abundant, information filtering becomes a key task for online users to obtain relevant suggestions. As most Websites can be ranked according to users' rating and preferences, relevance to queries, and recency, how to extract the most relevant item from the over-abundant information is always a key topic for researchers in various fields. In this paper, we adopt tools used to analyze complex networks to evaluate user reputation and item quality. In our proposed accumulative time-based ranking (ATR) algorithm, we incorporate two behavioral weighting factors which are updated when users select or rate items, to reflect the evolution of user reputation and item quality over time. We showed that our algorithm outperforms state-of-the-art ranking algorithms in terms of precision and robustness on empirical datasets from various online retailers and the citation datasets among research publications.



\end{abstract}


\end{frontmatter}

\section{Introduction}

As online information networks are becoming more popular, people are getting more information from the Web. Nevertheless, the Web is not useful if one cannot extract the relatively small amount of useful results out of a large pool of complicated information. Therefore, how to effectively filter information has become an increasingly important task in the online science community~\citep{liu2014information,albert2002statistical,xu2018optimal,wang2017new,jain2018movie,nie2015information}. In this case, one may rely on information acquired through interactions with strangers, which is useful for users and sometimes even profitable for the hosts. For instance, since we cannot guarantee the products in online retailers such as flikart.com and Amazon are of high quality as they claim, one can rely on the most conventional recommendation mechanism, i.e. the word of mouth; but instead of a face-to-face interaction, it now takes the form of an online reputation system in the cyberspace~\citep{bedi2018trust,qian2019three,ricci2015recommender,sun2015recommender,yin2018joint}. The information used by the recommendation mechanism is the interactive information between the user and the item, and the most suitable item is recommended to the user. Reputation mechanism provides a criterion to the problem of trust-building.~\cite{dellarocas20021}
These online reputation systems are effectively virtual word-of-mouth networks where users share opinions and scores on products such as companies, hotels, video contents and more. As a result, many online commercial Websites such as eBay, Amazon, and Netflix have introduced a rating system~\citep{dellarocas2000immunizing,goldberg1992using} for users to evaluate items, articulating accumulative score to reflect the quality of product.

In view of the high commercial values, corporations pay attention to their reputation and the reputation of their products on the Web and social networks. Reputation is thus considered one of the most important strategic resources~\citep{behrendt2015reputation,massucci2019measuring,nunes2019explaining,syed2018enterprise,taeuscher2019reputation,wenpan2019evolution,wu2018effect}. A good reputation can improve the corporate branding and its position in the market~\citep{centeno2018textual,ding2017rank,ponzi2011reptrak}. On the other hand, a bad reputation affects their business. Other than reputation, corporations also realize that online conversations and posts embed a continuous stream of valuable information~\citep{ren2018rigorous,roberts2012leveraging}. Companies that effectively analyze these information may obtain clues for innovation and improvement. Therefore, it is beneficial for corporations to track and improve their reputation on online platforms~\citep{liu2015ranking,sarstedt2013measuring}.

The quality of an item is closely related to the reputation of users. For example, some malicious users often give low scores and water army often gets high scores. Although the benefits gained from good reputation are well known, how to accurately quantify is uncertain. The most straightforward way is to consider the average rating by users. However, this method is sensitive to noisy information and malicious operations, and various alternative approaches are proposed. For instance, with BiRank~\citep{he2017birank} one iteratively assigns scores to vertices on a bipartite graph of users and products and finally get a stationary rank among items. Nevertheless, some users may give unreasonable scores because they consider that ratings are unimportant, or they are not familiar with the field of the items.

Other than average rating, another representative method is the iterative refinement (IR) method~\citep{laureti2006information}. In this method, user reputation is considered to be inversely proportional to the difference between his/her rating on an item and its estimated quality, i.e., the weighted average score of the item by all users. Item quality and user reputation are then iteratively updated until they become convergent. In addition, a new iterative optimization algorithm called correlation-based ranking (CR) algorithm~\citep{zhou2011robust} is obtained from optimizing IR by assigning a single reputation value for each scoring event, where user reputation is calculated by Pearson correlation between his/her assigned scores and the estimated quality of items~\citep{lawrence1989concordance,nagelkerke1991note}. This method is shown to be very effective in dealing with different malicious behaviors of spammers. In order to better cope with the malicious behavior and to enhance robustness, a reputation redistribution process is used to improve the reputation of well-known users and two penalty factors are applied to make the algorithm robust against the malicious behavior, which is called the iterative algorithm with reputation redistribution (IARR)~\citep{liao2014ranking}.

In this paper, based on the findings of related iterative algorithms,  we introduce the accumulative time-based ranking (ATR) algorithm, with two behavioral weight factors and two new iterative components introduced based on the time of the scoring events and the corresponding user and item degree at that time. The behavioral weight factor can significantly up-weigh the importance of user in a certain period of time, and we will show the proposed two-iterative-component accumulation process can effectively evaluate user reputation and item quality.

\section{Related works}
\subsection{Average score}
In online retailers, the rating on a product is usually calculated by averaging its score evaluated by users who have experience with the product, and is continuously updated when new scoring events occur; future potential buyers may favor products with higher average scores. Mathematically, the average score $\overline{r}_\alpha $ of item $\alpha$ is given as follows:
\begin{equation}
    \overline{r}_{\alpha} = \frac{\sum_{i\in \mathcal{U}_{\alpha}}r_{i\alpha}}{k_{\alpha}}
\end{equation}
where $\mathcal{U}_{\alpha}$ is the set of users rated item $\alpha$, $r_{i\alpha}$ denotes the cumulative effect of user $i$ on item $\alpha$, and $k_\alpha$ is the number of users who have collected item $\alpha$. The average score $\overline{r}_{\alpha}$ is used as the basis for ranking items. However, average scores have a large defect. The popularity of an item may fluctuate over time, and malicious rating by spammers may greatly influence the average score.

\subsection{Iterative refinement (IR)}

The iterative refinement (IR) algorithm~\citep{laureti2006information} considers user reputation as inversely proportional to the average squared error between his/her vector of item rating and the vector of item rating averaged over users. Nevertheless, the correlation-based ranking (CR) algorithm~\citep{zhou2011robust} to be introduced below is shown to be more robust against spamming ratings than the IR method and thus lead to a more accurate estimation of item quality.

\subsection{Correlation-based Ranking (CR)}

In the CR iterative algorithm proposed by Zhou et al~\citep{zhou2011robust}, it evaluates the quality $Q_{\alpha}$ of an item $\alpha$ as:
\begin{equation}
    Q_{\alpha} = \frac{\sum_{i\in \mathcal{U}_{\alpha}} R_{i} \cdot r_{i\alpha}}{\sum_{i\in \mathcal{U}_{\alpha}}R_{i}}
\end{equation}
where $\mathcal{U}_{\alpha}$ is the set of users rated item $\alpha$, $R_{i}$ is the reputation of user $i$ and $r_{i\alpha}$ is the rating of user $i$ on item $\alpha$. Every user's reputation is initially assigned according to his/her degree as $R_{i} = k_{i} / \left|\mathcal{O} \right|$, where $k_{i}$ is the degree of user $i$ and $\left|\mathcal{O} \right|$ represents the number of items in the set $\mathcal{O}$.

The algorithm calculates the correlation corr$_i$ of user $i$ between his/her assigned scores and the quality of items as follows:

\begin{equation}
    corr_{i} = \frac{1}{k_{i}} \sum_{\alpha \in N_{i}}\left(\frac{r_{i\alpha}-\overline{r}_i}{\sigma_{i}}\right)\left(\frac{Q_{i\alpha}-\overline{Q}_{\alpha}}{\sigma_{\alpha}}\right)
\end{equation}
where $k_i$ is the degree of user $i$, i.e. the number of item collected by user $i$, $N_i$ is the set of items collected by user i, $r_{i\alpha}$ is the score rated by user $i$ on item $\alpha$, $\overline{r}_i$ and $\sigma_i$ are the average rating by user $i$ and the standard deviation among ratings given by user $i$ respectively; similarly, $Q_{i\alpha}$ is the quality of item $\alpha$ evaluated by user $i$, $\overline{Q}_\alpha$ is the average quality of item $\alpha$ and $\sigma_\alpha$ is the standard deviation of the quality of item $\alpha$ evaluated by all collector users. The reputation of users is then assigned according to the value of corr$_i$:
when the value of corr$_{i}<0$ , the reputation of the user $R_{i}$ is 0. Conversely, if the corr$_{i}\ge 0$, the reputation of user $i$ is $R_{i}$ = corr$_{i}$.  In~\cite{zhou2011robust}, the ranking algorithm based on Pearson correlation coefficient is proved to be more powerful in dealing with spammers and able to evaluate the quality of objects more accurately.

\subsection{BiRank}

BiRank algorithm~\citep{he2017birank} ranks items on user-item bipartite graphs through iterations until convergence. BiRank is analyzed by graph regularization, and a complementary Bayesian view is proposed. Firstly, the ranking vector is randomly initialized. Then, the iteration process is executed until convergence. During the iteration process, the BiRank algorithm utilize  query vectors, i.e. sparse vectors with the entries for the collected items to be non-zero, thus representing vectors encoding user preference, to guide the ranking process. The iteration can be expressed as follows:

\begin{equation}
\begin{array}{l}{\mathbf{v}=\alpha S^{T} \mathbf{u}+(1-\alpha) \mathbf{v}^{0}} \\ {\mathbf{u}=\beta S \mathbf{v}+(1-\beta) \mathbf{u}^{0}}\end{array}
\label{BiRank}
\end{equation}

In Eq.~(\ref{BiRank}), ${\mathbf{v}^{0}}$ and ${\mathbf{u}^{0}}$ represent query vectors, ${\alpha}$ and ${\beta}$ are hyper-parameters which weigh the importance of the previous query vectors. After convergence of the query vectors ${\mathbf{v}}$ and ${\mathbf{u}}$, they become vectors for ranking items. Besides, one can express $S=D_{u}^{-\frac{1}{2}} W D_{v}^{-\frac{1}{2}}$, where $D_u$ and $D_v$ are diagonal matrices and represent the weighted degrees of all vertices in $\mathcal{U}$ and $\mathcal{V}$, i.e. the sum of weights on connected edges with $\mathcal{U}$ and $\mathcal{V}$ to be the sets of users and items respectively. As such, $W$ is the matrix of which entries are edge weights of the graph as a $\left| \mathcal{U} \right| \times \left| \mathcal{V} \right|$~\citep{he2017birank}.

\subsection{The iterative algorithm with reputation
redistribution(IARR)}

Based on the CR algorithm~\citep{zhou2011robust}, Liao et al.~\citep{liao2014ranking} used the reputation calculated by the CR algorithm as the temporary reputation $TR_{i}$ of user $i$, which is used to redistribute user reputation and obtain a new reputation $R_i$ for user $i$. The algorithm is called the iterative algorithm with reputation
redistribution (IARR) algorithm. IARR is based on the CR algorithm but has the reputation redistribution in order to filter noisy information during the iterations, so as to improve the accuracy in items' quality ranking.~\cite{liao2014ranking} $TR_{i}$ in IAAR is same with $corr_{i}$ in CR, representing the similarity between ratings by users and weighted average rating. Differently, the reputation $R_i$ of user $i$ in IARR is not equal to $TR_{i}$ but caculated as follows:

\begin{equation}
    TR_{i} = \frac{1}{k_{i}} \sum_{\alpha \in \mathcal{O}_{i}}\left(\frac{r_{i\alpha}-\overline{r}_i}{\sigma_{r_{i}}}\right)\left(\frac{Q_{i\alpha}-\overline{Q}_\alpha}{\sigma_{\alpha}}\right)
\end{equation}

\begin{equation}
    R_{i} = TR_{i}^{\varphi}\cdot\frac{\sum_{i\in \mathcal{U}}TR_{i}}{\sum_{i\in \mathcal{U}}TR_{i}^{\varphi}}
\end{equation}

where $\varphi$ is a parameter of the algorithm, and the reputation of users is calculated by adjusting the value of the parameter. The purpose to redistribute user reputation by the IARR algorithm is to further up-weigh users with high reputation, since evaluation by users with high-reputation are more reliable, and can reduce the interference by malicious user behavior~\citep{jnanamurthy2013detection}. During the iterations, it can filter the noise by diminishing the weight of users with low $TR_{i}$. With the accumulation of effect in each iteration, the results in the accuracy of item quality will get a large improvement.

\section{Methods and data}


\subsection{Bipartite networks}

The bipartite network is a mathematical representation the real world relations. It is named bipartite as there are two types of nodes. For instance, if we want to represent the relationship between users and products in an online shop, one type of nodes would be user, while the other would be products. A link between a user and a product could mean that the user has bought the product. Similarly, networks can be created for users and posts in online social media, or users and music on music Websites. Mathematically, the network G = $\langle$E, V$\rangle$ contains the set of nodes $V$ and the set of edges connecting the nodes $E$. If the set $V$ of nodes can be divided into two subsets $X$ and $Y$ with no connections within the subset, the network is said to be bipartite.

Most social networks can be represented as bipartite networks. However, it is not always the case. For instance, the citations networks of scientific papers is usually studied as a relationship between papers, rather than between the scientists who write the papers. One can employ recommendation algorithms conventionally applied on bipartite networks to represent the complex citation networks between papers, though one cannot identify two subsets of papers and their references as every publication can be cited in the future~\citep{bao2015recommendations,gao2015content,gualdi2011tracing}. The algorithms developed for bipartite networks can be applied on monopartite networks by defining a set $A$, and then defining a set $B$, identical to set $A$. In the citation network, an edge between sets $A$ and $B$ corresponds to a citation from a paper in set $A$ to a paper in set $B$, which is a peer-to-peer network.

\subsection{Accumulative time-based ranking(ATR)}
In this paper, we develop an approach to analyze the interactions between users in social networks, and produce a ranking of users and products. This type of ranking can be for instance used to make recommendation~\citep{lu2012recommender}, or to predict the Oscar nominations~\citep{pardoe2008applying}.

Compared to the existing reputation evaluation algorithms, our proposed ATR method focuses on combining the iterative approach to rank user reputation and item quality with the temporal dependence.
The main idea of this algorithm is to take into account the whole history of the network to compute the reputation of users and the quality of items. Similar to previous methods, the ratings by users with higher reputation weight more in the determination of the quality of items~\cite{liao2014ranking}.

We denote the quality of item $\alpha$ by $Q_{\alpha}$, and the reputation of user $i$ by $R_{i}$. The iterative process of the ATR algorithm uses the aggregated reputation of users and the aggregated quality of items to improve their weighting in the algorithm. Consequently, the value of user reputation and item quality are continuously updated in the iterative process until convergence. The notations of the algorithm are summarized in Table~\ref{notation}.

\begin{table}[h]
\setlength{\abovecaptionskip}{0.cm}
\setlength{\belowcaptionskip}{-0.cm}
\caption{Notation table and description}
\resizebox{\textwidth}{!}{%
\begin{tabular}{cccc}
\hline
\textbf{Notation table}                  & \textbf{Description}                                                           & \textbf{Notation table}             & \textbf{Description}                                              \\ \hline
\textbf{$\mathcal{U}$}                                        & The set of users
& \textbf{$\left|\mathcal{U} \right|$}                          & The number of all users  \\
\textbf{$\mathcal{O}$}                                        & The set of items (such as music, movies, etc.)
& \textbf{$\left|\mathcal{O} \right|$}                                        & The number of all items \\
\textbf{$\mathcal{U}_\alpha(t)$}                                        & The set of users collected item $\alpha$ in year $t$
& \textbf{$\mathcal{O}_i(t)$}                          & The set of items collected by user i in year $t$  \\
\textbf{$\mathcal{U}(t)$}                                        & The set of users in year $t$
& \textbf{$\mathcal{O}(t)$}                          & The set of items in year $t$  \\
\textbf{$\overline{r}_i$} & The average rating on items given by user $i$  &\textbf{$r_{i\alpha}$}                            & The average rating (score, like, etc.) of item $\alpha$ rated by users\\                 \textbf{$\sigma_{i}$}                            & The standard deviation of ratings given by user $i$     & \textbf{$k_{i}(t)$}                                 & The degree of user $i$ in year $t$\\                                                      \textbf{$\tilde{\sigma}_{\alpha}$}                       & The standard deviation of the quality of item $\alpha$
& \textbf{$\tilde{k}_{\alpha}(t)$}                            & The degree of item $\alpha$ in year $t$\\                               \textbf{$N(t)$}                             & The total number of user ratings evaluation in year $t$
& \textbf{$Q_{\alpha}(t)$}                             & The quality of item $\alpha$ in year $t$\\                              \textbf{$\overline{Q}_{\alpha}(t)$}                            & The average quality value of the item $\alpha$ in year $t$
& \textbf{$R_{i}(t)$}                                  & The reputation of user $i$ in year $t$\\                       \textbf{$W_{i}(t)$}                           & The weight of the user $i$ in year $t$
&\textbf{$\mathcal{Y}$}                           & The set of year in each individual dataset    \\                                \textbf{$\tilde{W}_{\alpha}(t)$}                      & The weight of the item $\alpha$ in year $t$   & \textbf{$\left|R \right|$}                          & The number of all ratings  \\
\textbf{$\overline{K}_\mathcal{U}$}                      & The average degree of all users   & \textbf{$\overline{K}_\mathcal{O}$}                          & The average degree of all items
\\
\textbf{$M$}                          & The matching numbers of getting Oscar movies or Nobel Prize   &  \textbf{$f$}                      & The top percent of the list of items ranked by quality
 \\
\textbf{$N$}                          & The total comparisons counts in experimental evaluation metrics   &  \textbf{$N_{sam}$}                      & The number of samples when computing $RMSE$
 \\
\textbf{$AUC^{real}$}                          & The value of AUC obtained by the original datasets without malicious behavior  & \textbf{$AUC^{ran}$}                          & The value of AUC obtained by the datasets with random ratings
\\
\textbf{$n$}                          & The number of users assigning random ratings
\\ \hline
\label{notation}
\end{tabular}%
}
\end{table}

We use the following initial configuration for better convergence in the iterative procedures. The first assumption in our algorithm is that the initial reputation of users and the initial quality of items are dependent on the number of ratings they assign or receive. The ratings of users and items with higher weights depend on high weight in user-item evaluation.
On the other hand, reputation and quality in IARR\cite{liao2014ranking} are not considered with temporal dependency and are vulnerable to malicious behaviors. Such approaches do not take into account the history of users and items during evaluation, for instance, users which are active over multiple years are less likely to be spammers. In the case of ATR, we thus define the weight of a user by Eq. (\ref{weight_u}) and the weight of an item by Eq. (\ref{weight_i}):

\begin{equation}
W_{i}(t) =\frac{k_{i}(t)}{N(t)}.
\label{weight_u}
\end{equation}
\begin{equation}
\tilde{W}_{\alpha}(t) =\frac{\tilde{k}_{\alpha}(t)}{N(t)}.
\label{weight_i}
\end{equation}
where $N(t)$ is the total number of user ratings evaluation in year $t$.

The rating of user $i$ on item $\alpha$ is expressed as $r_{i\alpha}$. The reputation of users and the quality of items is initialized as:
\begin{equation}
R_{i} =\frac{\sum_{\alpha \in \mathcal{O}_{i}(t), t\in \mathcal{Y}}r_{i\alpha}\cdot W_{i}(t)}{\sum_{t\in \mathcal{Y}} k_{i}(t)}
\label{r_1}
\end{equation}
\begin{equation}
Q_{\alpha} =\frac{\sum_{i\in \mathcal{U}_{\alpha}(t), t\in \mathcal{Y}}r_{i\alpha}\cdot R_{i}(t)\cdot \tilde{W}_{\alpha}(t)}{\sum_{i\in \mathcal{U}_{\alpha}(t), t\in \mathcal{Y}}R_{i}(t)\cdot \tilde{W}_{\alpha}(t)},
\label{q_1}
\end{equation}
where $\mathcal{O}_i(t)$ and $\mathcal{U}_\alpha(t)$ are the sets of objects evaluated by user $i$ in year $t$, and the set of users who evaluated item $\alpha$ in year $t$, respectively. $\mathcal{Y}$ is the set of years considered for the evaluation.

The main rationale given by Eq. (\ref{weight_u}) - (\ref{q_1}) can be described as follows: the reputation of a user depends on the total number of ratings the user assigns in a period of time, and the quality of an item depends on the frequency of it being rated by users in a period of time. For example, the reputation of users in Amazon depends on the number of movies he/she watched and rated in a period of time. The more movies the user watched, the greater his/her reputation. On the item side, the quality of items depends on the number of ratings it received in a period of time. If a lot of users watched and rated the movie in a period of time, the movie is of a higher timely quality, and vice versa.

The next crucial component of the ATR method is aggregation. The reputation of users is not established at a single time instance, but instead through a process of accumulation over the years. The reputation of a user is thus not defined in a single point of time, but by the succession of his/her behaviour over the years.

The iterative equations to determine the reputation of users and quality of items writes

\begin{equation}
\mathrm{accu}_{i}(t) = \frac{R_{i}(t-1)}{\sqrt{k_{i}(t)}}\cdot \sum_{\alpha\in \mathcal{O}_{i}(t)}\frac{\tilde{k}_{\alpha}(t)}{Q_{\alpha}(t-1)\cdot \sqrt{r_{i\alpha}}}
\end{equation}

\begin{equation}
\mathrm{accu}_{\alpha}(t) = \sum_{i\in \mathcal{U}_{\alpha}(t)}\frac{R_{i}(t-1)\cdot Q_{\alpha}(t-1)}{\tilde{k}_{\alpha}(t)}\cdot (1-\frac{1}{\sqrt{{k}_{i}(t)}})
\end{equation}

\begin{equation}
R_{i}(t) = \frac{1}{\mathrm{accu}_{i}(t)\cdot max\left\{logk_{i}(t)\right\}}\sum_{\alpha\in \mathcal{O}_{i}(t)}\left(\frac{W_{i}(t)\cdot (r_{i\alpha}-\overline{r}_i)}{\sigma_{i}}\right)\cdot\left(\frac{\tilde{W}_{\alpha }(t)\cdot(Q_{\alpha}(t-1)-\overline{Q}_\alpha(t-1))}{\sqrt{\frac{\tilde{\sigma}_{\alpha}}{k_{i}(t)}}}\right)
\end{equation}

\begin{equation}
Q_{\alpha}(t) = \frac{\sum_{i\in \mathcal{U}_{\alpha}(t)}(\tilde{W}_{\alpha}(t)\cdot r_{i\alpha})}{\sum_{t \in \mathcal{Y}}(\tilde{W}_{\alpha}(t))}\cdot \mathrm{accu}_{\alpha}(t)\cdot \tilde{k}_{\alpha}(t)
\end{equation}
We name $\mathrm{accu}_i(t)$ and $\mathrm{accu}_\alpha(t)$ the behavioral weight factors.
The reciprocals of degree $k_i(t)$ and $r_{i\alpha}$ in Eq. (11) are to penalize potentially malicious behavior, for instance, a large number of ratings given by a specific user $i$, i.e. large $k_i(t)$, and a high rating value $r_{i\alpha}$ in the specific time period $t$. Nevertheless, not to penalize large but normal values of $k_i(t)$ and $r_{i\alpha}$, we used the square root values $\sqrt{k_i(t)}$ and $\sqrt{r_{i\alpha}}$ in the equations. We propose the square root to study behavioral dynamics based on the fluid dynamics theory derived from~\cite{jiang2014fluidrating}.

The quality of items calculated iteratively by the behavior weight factor, a single malicious action will not have much effect on results. They can reduce the impact of malicious behaviour. The user reputation is computed as the Pearson correlation coefficient between the rating vector of user and the quality vector of the corresponding items as the temporal reputation. The process runs iteratively until $|Q_{\alpha}(t)-\overline{Q}_{\alpha}(t)| < r$, where $r$ is the threshold, set to $r=10^{-4}$ in this work. When $r=10^{-4}$, the quality of items doesn't change and the iteration becomes stable.

We note that user reputation depends on the number of times he/she rated items in a period of time, and so as the quality of items. Therefore, by taking the reciprocal of the number of ratings, the influence of malicious behavior on the item quality is reduced. More details of our ATR method implementation and datasets are available on the Baidu Netdisk \footnote{Link: https://pan.baidu.com/s/1J0K9UULYCjZGk7yA7Qlvow
Access code: ux3t}.

\subsection{Experiment data}

\begin{table}[h]
\setlength{\abovecaptionskip}{0.cm}
\setlength{\belowcaptionskip}{-0.cm}
\caption{Datasets information}
\setlength{\tabcolsep}{7mm}{%
\centering
\begin{tabular}{cccccc}
\hline
\textbf{Dataset}   & \textbf{$\left| \mathcal{U} \right|$} & \textbf{$\left| \mathcal{O} \right|$} & \textbf{$\overline{K}_\mathcal{U}$} & \textbf{$\overline{K}_\mathcal{O}$} & \textbf{$\left| R \right|$} \\ \hline
\textbf{Artificial Network}    & 6000                & 4000                  & 81                                        & 126                                          & 480000               \\
\textbf{Amazon}    & 16834                & 26258                  & 30                                        & 19                                          & 4893777               \\
\textbf{APS}       & 449935               & 449935                 & 12                                        & 11                                          & 4672812               \\
\textbf{Movielens} & 10702                & 19931                  & 509                                       & 306                                         & 6099708               \\
\textbf{Netflix}   & 25000                & 17734                  & 207                                       & 292                                         & 1070556               \\ \hline
\end{tabular}%
}
\end{table}

\begin{figure}[htb]
\setlength{\abovecaptionskip}{0.cm}
\setlength{\belowcaptionskip}{-0.cm}
  \centering
  \includegraphics[scale=0.6]{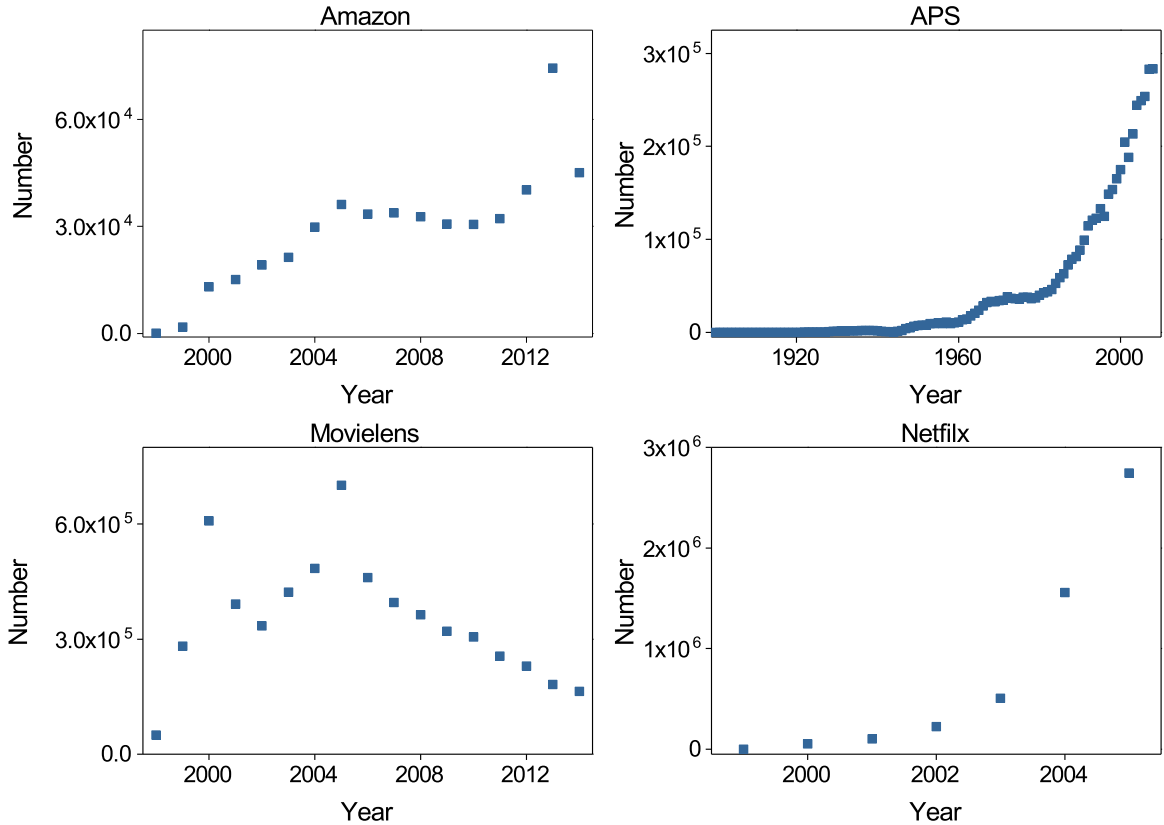}
\caption{The number of rating records in the four real datasets over years.}

\end{figure}

\subsubsection{Artificial networks}
In order to validate the effectiveness of our proposed method in a more general way, we generate a artificial network with $\left| \mathcal{U} \right|= 6000$ and $\left| \mathcal{O} \right| = 4000$. The rating network have a certain sparsity ($\beta = 0.2$) and we will add each rating to the network one by one. Therefore, the artificial network will finally have $\left| \mathcal{U}\right| \left| \mathcal{O}\right| * \beta = 4.8 * 10^{5}$ links. The construction process of artificial networks is widely used in previous research works~\cite{zhou2011robust,liao2014ranking,vidmer2015unbiased}.




\subsubsection{Real networks}
In this paper, we verify our proposed ATR algorithm by four real datasets, which are obtained on Amazon, APS, Movielens and Netflix.

(1) Amazon Dataset: Amazon is the largest company in the US online retailing industry, covering areas such as books, movies and more\footnote{https://www.amazon.com/}. Here we used a small dataset in Amazon, consisting of 16843 users and 26,258 movies. The value of ratings on the movie is from 1 to 5, which spans from year 1997 to 2014.

(2) APS dataset: The American Physical Society (APS) is the second largest physics organization in the world\footnote{https://journals.aps.org}. The APS dataset studied here is a citation network between papers, including the citation relationship between 449,935 articles. The articles we analysed are published from year 1893 to 2009. In the APS network, one can consider papers as users and items as articles cited by other papers.

(3) MovieLens dataset: MovieLens is one of the oldest recommendation systems\footnote{https://www.grouplens.org}. It is a non-commercial, research-oriented experimental online platform. We used a subset of the complete data. In this subset, there are 5 million rating records, each with a score from 1 to 5.

(4) Netflix dataset: Netflix is an American company originally mainly engaged in the online rental business of customized DVDs and high-quality compact discs\footnote{https://www.netflixprize.com}. By extracting a small dataset from the data provided by it, 25,000 users are selected and each user rated at least 20 movies on average. Then, all the movies that have been rated by users have a score from 1 to 5.


\subsection{Experimental evaluation metrics}

In order to analyse the capability of the algorithm, a wide range of accuracy evaluation metrics are used including AUC, precision, recall and the F value~\citep{davis2006relationship,hanczar2010small}. In the Amazon dataset, the Movielens and the Netflix datasets, items are movies, while in the APS dataset, items are research publications:

(1) AUC: AUC is a standard metric used to measure the accuracy of classification or prediction tools. To compute AUC, we make $N$ independent comparison between the score of the correct items identified by the algorithm with that of the incorrect items. Among the $N$ comparisons, if the correct items have a higher score than the incorrect items in $N_1$ of the comparisons, while in $N_2$ of the comparisons the correct and the incorrect items have the same score, then the value of AUC is given by AUC = $(N_1 + 0.5N_2) / N$. If all items are ranked randomly, AUC = 0.5; if the correct items always have a higher score than the incorrect items, AUC = 1.

(2) Precision: Precision measures the accuracy of correct retrieval. Precision is the ratio between the number of correctly retrieved items to the total number of retrieved items. Suppose we let $X$ to be the set of correct items, and $Y$ to be the set of items identified by the algorithm, then the value of Precision is given by $\left| X \cap Y \right| / \left| Y \right|$.

(3) Recall: Recall also measures the accuracy of correct retrieval, and is the ratio between the number of correctly retrieved items and the total number of correct items. In other words, Recall = $\left| X \cap Y \right| / \left|  X \right|$.

(4) $F$ value: To quantify the accuracy for correct retrieval, recall and precision: when the value of recall is high, the precision is low, and when the precision is high, the recall is low. Therefore, the $F$ value based on precision and recall is used for a comprehensive measurement: $F$ = 2 * Precision * Recall / ( Precision + Recall ).

\section{Results}

\begin{figure}[htb]
\setlength{\abovecaptionskip}{0.cm}
\setlength{\belowcaptionskip}{-0.cm}
  \centering
  \includegraphics[scale=0.3]{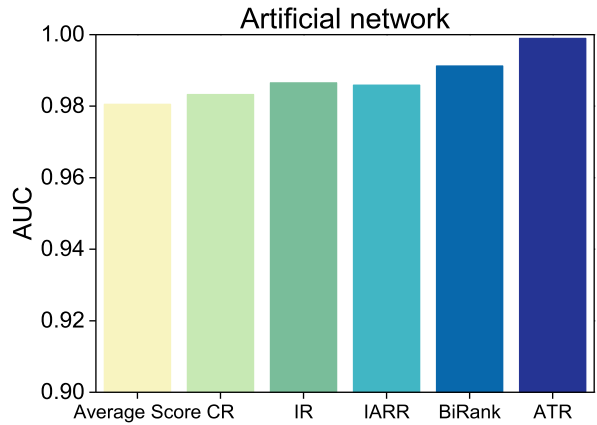}
\caption{The AUC of different algorithms in artificial network.}

\end{figure}

In this experiment, the average score, CR~\citep{zhou2011robust}, IR~\citep{laureti2006information}, IARR~\citep{liao2014ranking}, BiRank~\citep{he2017birank} and ATR are verified on the five datasets. Since item quality and user reputation are accumulated over a long period of time, there is always some inaccuracy. For example, the quality of some commodities is initially good, but when time passes some of the stocks have lower qualities and received lower ratings, which make the ratings of good items lower. The same applies to user reputation.

Since the evaluation of item quality is not formed instantaneously, but instead resulted from a long-term accumulation process, which changes with user subjective and emotional choice and experience. For example, when a good product just hits the shelves, users are influenced by favorable comments towards the product and may end up choosing the product. As time goes on, the rating of the product gradually decreases. Therefore, the quality of an item and the reputation of a user are determined by the accumulation of the reputation of multiple users in a time period.

\begin{figure*}[htb]
  \centering
  \includegraphics[scale=0.6]{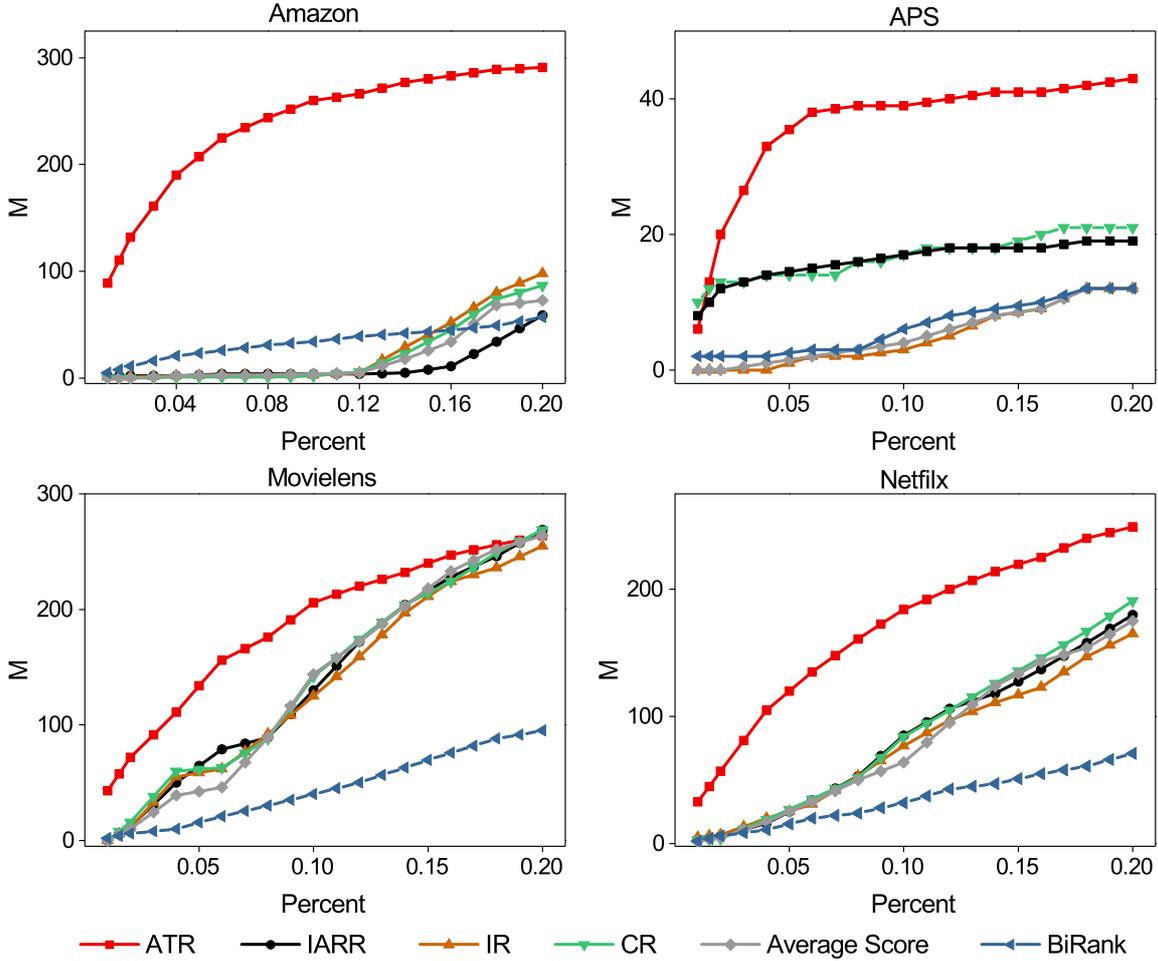}
\setlength{\abovecaptionskip}{0.cm}
\setlength{\belowcaptionskip}{-0.cm}
\caption{The relation between $M$ and the top percent of the list of items ranked according to their quality estimated by six algorithms.}
\label{m_by_percent}
\end{figure*}

\begin{figure*}[htb]
  \centering
  \includegraphics[scale=0.6]{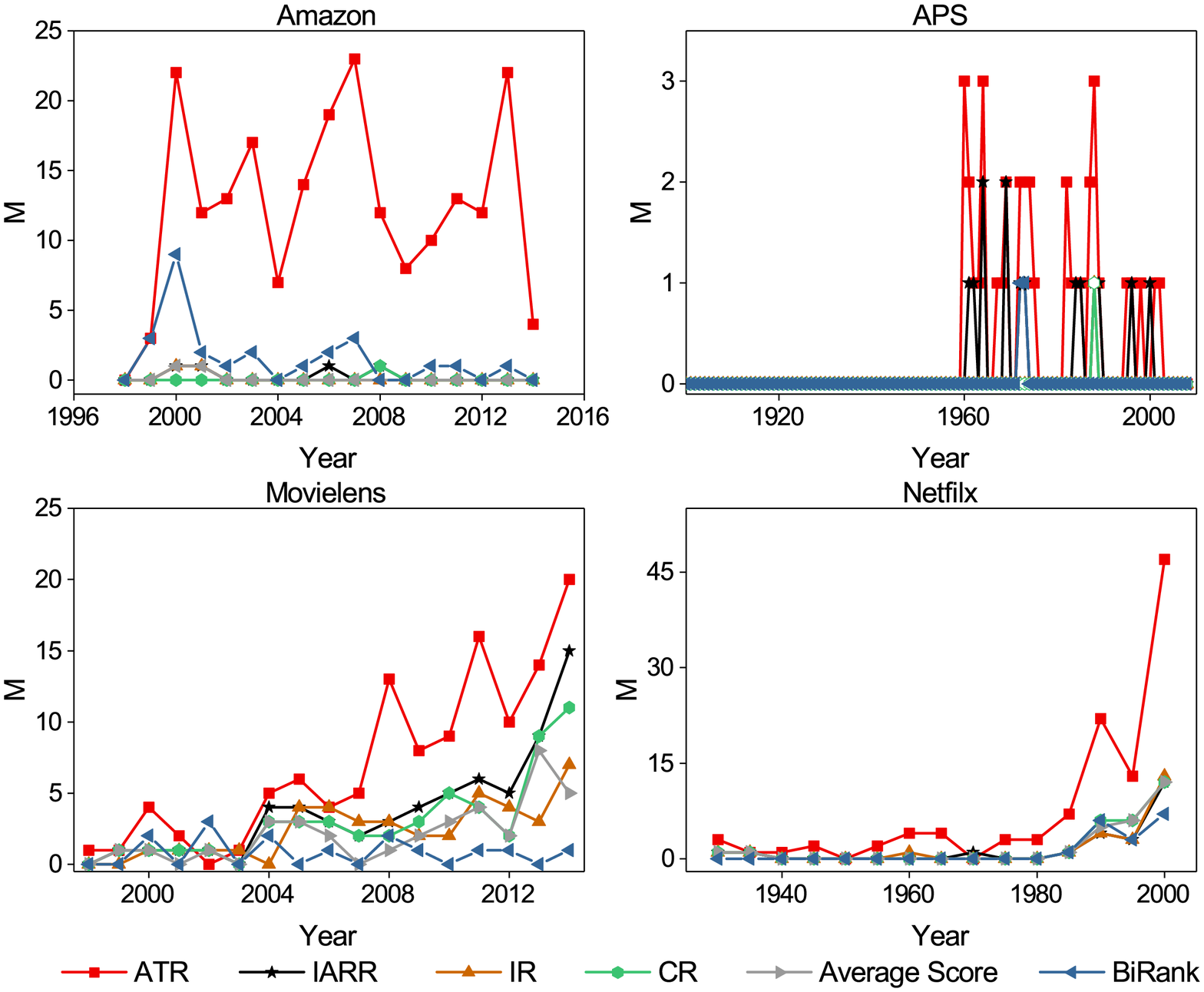}
\setlength{\abovecaptionskip}{0.cm}
\setlength{\belowcaptionskip}{-0.cm}
\caption{Dependence of M on the year in six algorithms, where $f=5$.}
\label{m_by_year}
\end{figure*}

\subsection{Identification of High-quality Movies and Publications}

We apply our proposed ATR algorithm and other benchmark algorithm to identify Oscar-winning movies by the ratings in Amazon, Movielens and Netflix datasets, and the Nobel Prize winning publications in the APS datasets.

Since the quality of items in the four empirical datasets is not known, we used awarded movies or publications as the target items to be identified by the algorithms. In the movie datasets including the Amazon, Movielens and Netflix datasets, we select 521 Oscar-winning films from 1928 to 2014 as the target items. We then apply various algorithms to rank the quality of movies, and denote the matching number of Oscar-winning movies in the list of the top $5\%$ ($f=5$) of estimated quality by a specific algorithm to be $M$. Similarly, we identify 87 Nobel Prize research papers in the APS dataset and denote the matching number of Nobel Prize winning publications in the top $5\%$ ($f=5$) of the quality list to be $M$.

\begin{figure*}[htb]
  \centering
  \includegraphics[scale=0.65]{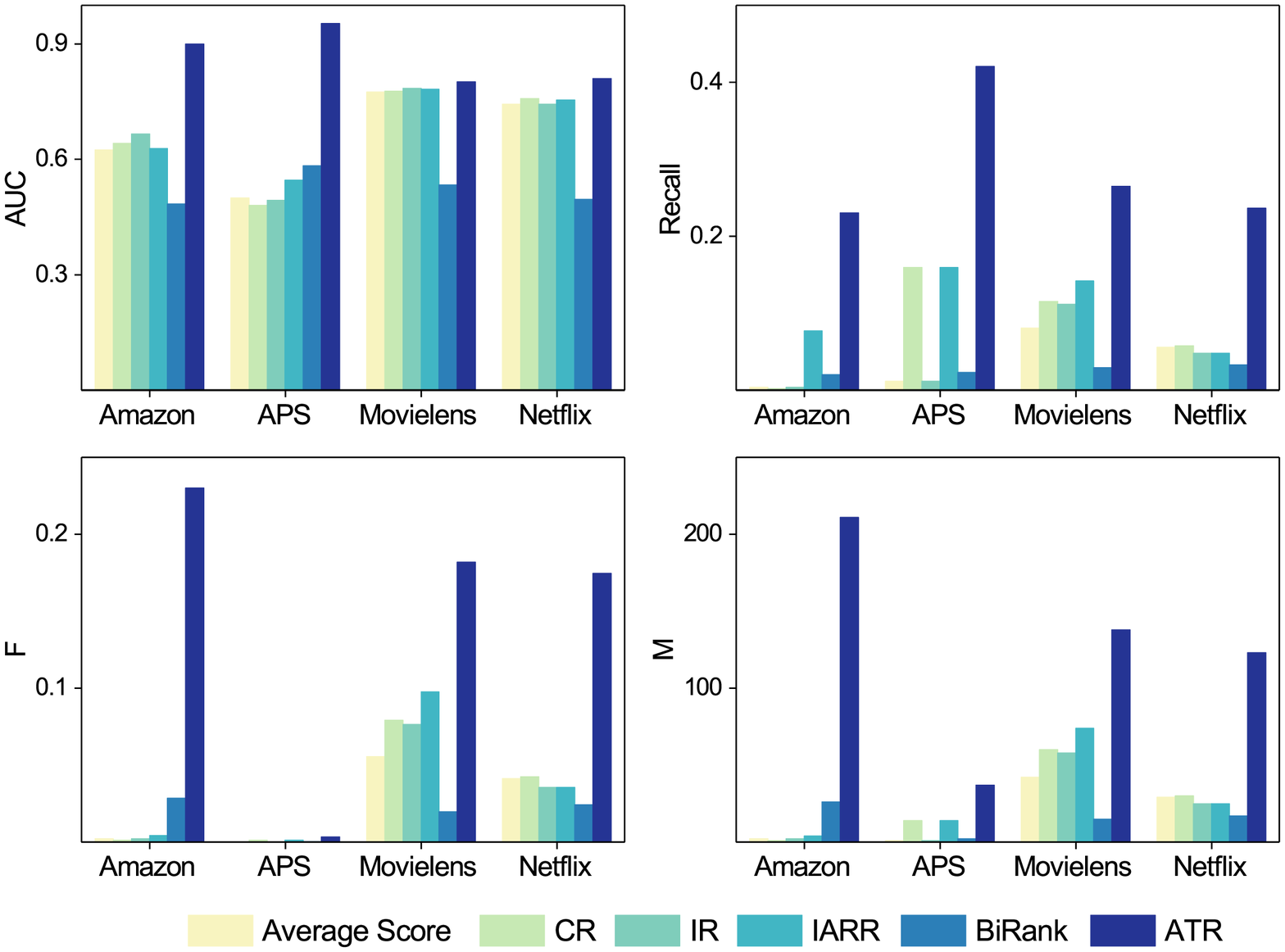}
\setlength{\abovecaptionskip}{0.cm}
\setlength{\belowcaptionskip}{-0.cm}
\caption{The results of AUC, recall, the $F$ and the $M$ values obtained by different algorithms.}
\label{metrics}
\end{figure*}

In Fig.~\ref{m_by_percent}, we examine items in the top $f\%$ of the item list as ranked by their quality estimated by the different algorithms. As we can see in Fig.~\ref{m_by_percent}, when $f$ increase, the values of $M$, i.e. the total number of Oscar-winning movies or Nobel Prize winning publications identified by all the algorithms tend to rise. As we can also see, all the benchmark algorithms can identify high-quality movies and publications, but the ATR algorithm always results in a larger value of $M$, i.e. outperforms other benchmark algorithms. This also shows that the ATR algorithm can well utilize the interactive procedures to update the evaluation on both user reputation and item quality, such that it can achieve better results.

\begin{figure*}[htb]
  \centering
  \includegraphics[scale=0.68]{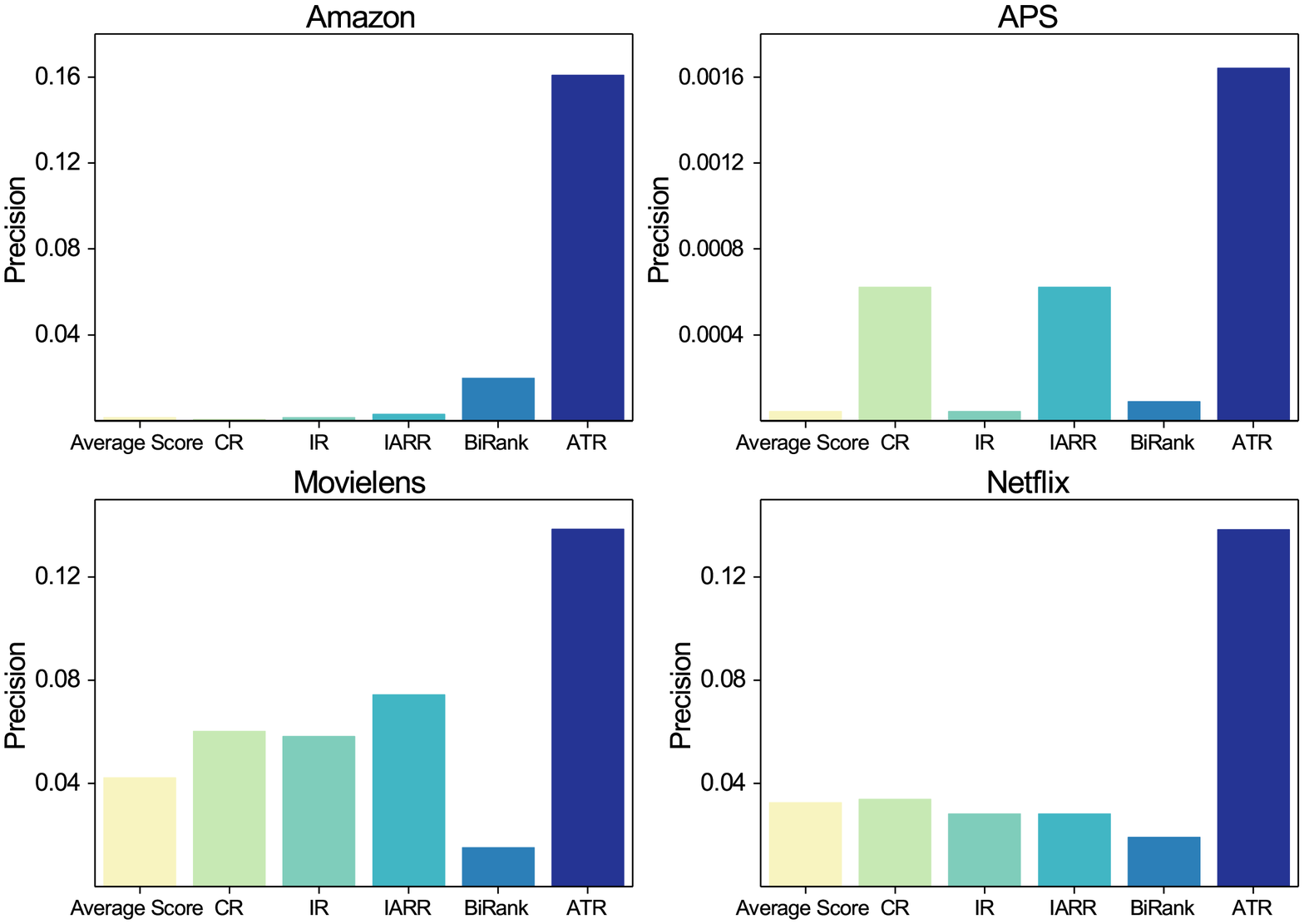}
\setlength{\abovecaptionskip}{0.cm}
\setlength{\belowcaptionskip}{-0.cm}
\caption{The results of precision in $M$ obtained by different algorithms.}
\label{pre}
\end{figure*}

\subsection{Matching numbers by years}

In Fig.~\ref{m_by_year}, we show that the results of $M$ obtained by various algorithms over time. As we can see, the ATR algorithm outperforms the other algorithms in the four datasets. First of all, the ATR algorithm obtains much larger values of $M$ in the Amazon dataset compared to the other algorithms. We also see that from our results, some of the award-winning movies identified by the other algorithms are overlapped with each other, suggesting that those algorithms can only identify obviously good movies. In addition, there are malicious behaviors or operations such as water army in the Amazon dataset~\citep{xu2013uncovering}, which leads to a reduction in accuracy obtained by the other benchmark algorithms. On the other hand, we see that other benchmark algorithms also show poor accuracy in the APS dataset. Finally, in the Netflix and Movielens dataset, it is clear that the ATR algorithm obtains larger values of $M$ and hence a higher accuracy compared with the other algorithms.

Other than the $M$, we also examine the standard metrics including AUC, precision, recall and the $F$ values. As we can see from Fig.~\ref{metrics} and Fig.~\ref{pre}, the ATR algorithm again outperforms other algorithms in terms of these metrics, and shows the better capability in identification of high-quality items. Specially, in Amazon and APS dataset, the values of precision obtained by the method of the average score and the IR are almost zero, suggesting that the reputation is not well evaluated. However, the ATR always obtains the largest value of precision in the four datasets. We also remark that the ATR algorithm has less computational cost on top of its better performance.

\begin{figure*}[htb]
\setlength{\abovecaptionskip}{0.cm}
\setlength{\belowcaptionskip}{-0.cm}
  \centering
  \includegraphics[scale=0.6]{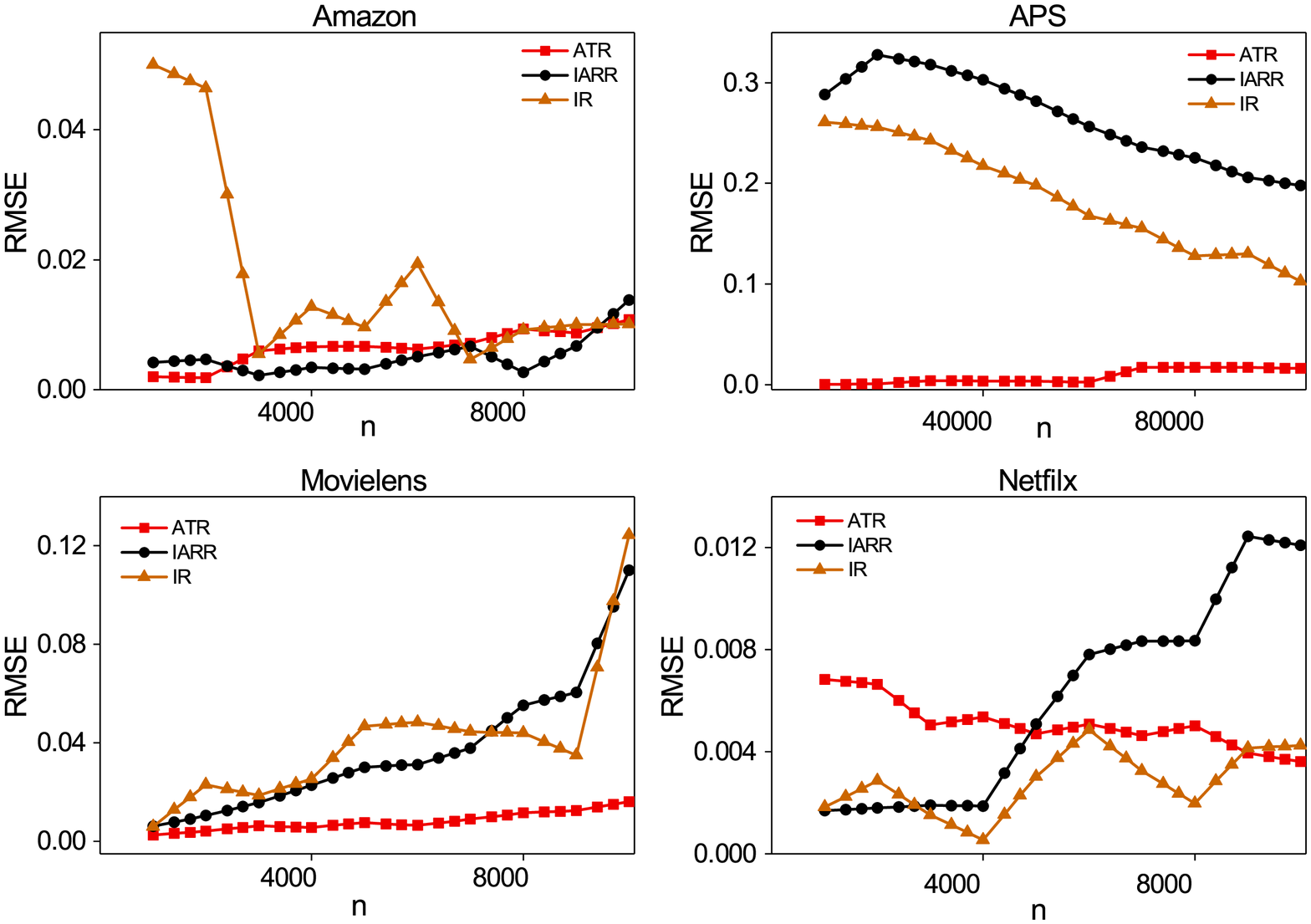}
\caption{The relation between RMSE and $n$, where $n$ represents the user numbers assigning random ratings.}
\label{rmse}
\end{figure*}

\subsection{Robustness}

Robustness refers to the ability of the algorithm to maintain a good result under various parameters, e.g. structure and size, and scenarios such as malicious attacks. Robustness could be used to evaluate the ability of the algorithm to reduce the influence by random ratings. We examined the robustness of our ATR algorithm and the other reputation evaluation algorithms through intentionally generating random ratings.

Initially, we assumed that there is no malicious behavior in all datasets we studied. In other words, the ratings by users on specific items are independently assigned according to their preferences. We then set $n$ from 1000, 2000, 3000 to 10000 users to assign random ratings in the Amazon dataset, and from 10000, 20000, 30000 to 100000 publications to have random citations in the APS dataset. We apply the same in the Movielens dataset and Neflix dataset as in the Amazon dataset. In the APS dataset, the rate of the references of papers in the original APS dataset is 5 (we can think that the reference papers cited by authors are usually highly relevant). 1-5 represents the author's subjective emotional tendency, i.e. the rating to a paper. The higher the ratings, the higher the quality of research as judged by that the authors citing the papers as references. We then compute the RMSE as shown in Eq.~(\ref{rmse_equ}):

\begin{equation}
    RMSE = \sqrt{\frac{\sum_{i=1}^{N_{\rm sam}}(AUC^{\rm ran}_{i}-AUC^{\rm real}_{i})^{2}}{N}}
\label{rmse_equ}
\end{equation}

where $AUC^{\rm ran}$ is the value of AUC obtained by the datasets with random ratings, $AUC^{\rm real}$ is the value of AUC obtained by the original datasets without malicious behavior, the label $i$ corresponds to the index of samples, and $N_{\rm sam}$ corresponds to the number of samples.

As we can see in Fig.~\ref{rmse}, the values of RMSE obtained by the IARR and ATR algorithms only increase slowly with the number of users or publications with random ratings or citations, suggesting that these two algorithms are more robust against malicious behavior. The overall robustness of the ATR algorithm is better than that of other benchmark algorithms, while the IR has poor anti-interference capability. It can also be seen from Fig.~\ref{rmse} that in the APS dataset, the ATR algorithm does not fluctuate much and tends to be stable as the number of random citations increase, indicating that the algorithm has good robustness. The IARR and IR algorithms show a declining trend, and the AUC gradually approaches the its values on the original datasets.

\section{Conclusion and discussion}
The relationship between the reputation of a product on the Web and its sales has been strengthening. However, the complicated and extensive amount of information on the Web may misguide user choices. In view of this situation, corporations recognize a higher importance to the improvement of service recommendation and user experience. In this paper, we express the relationship between users and items as bipartite graphs and introduce the accumulative time-based ranking (ATR) algorithm by taking into account the temporal factors of ratings to improve the evaluation of reputation found by the iterative algorithm with reputation redistribution (IARR).

On the basis of the ATR algorithm, its universality and robustness are explored. The temporal datasets from Amazon, Movielens, Netflix and the American Physical Society (APS) are selected for testing, and the results from the algorithm are evaluated by common evaluation indicators such as AUC, precision, the F value and recall value in identifying award-winning movies or publications in the datasets. At the same time, the robustness of the algorithm against random ratings is also explored. Remarkably, our purposed ATR reputation evaluation algorithm largely improves the accuracy of reputation evaluation and robustness against random ratings when compared to the other benchmark algorithms. However, although our algorithm shows a good performance, user's malicious behavior such as the water army in reality may affect the reputation evaluation. Therefore, the application of big data for reputation evaluation may be the next goal.

In summary, in this paper we introduced aggregating factors to evaluate reputation and quality in the reputation evaluation algorithm. However, malicious behavior may also have great impacts on reputation or quality aggregation. In addition, the attributes of items such as movie genre and cast are not considered in our algorithm. For instance, a celebrity may impact significantly how users select a movie. Potential future work is to study in-depth reputation and quality evaluation from the above perspectives.

\section*{Competing interests}

The authors declare that they have no competing interests.

\section*{Author's contribution}
The work presented in this paper corresponds to a collaborative development by all authors. Conceptualization, H.L, C.H.Y, Y-C.Z and Q-X.L; Data Curation, Q-X.L and Z-C.H; Formal Analysis, H.L, Q-X.L and C.H.Y; Funding~Acquisition, H.L and C.H.Y; Methodology, H.L, C.H.Y and Y-C.Z; Resources, H.L; Software, Q-X.L and Z-C.H; Writing---Original Draft, H.L, Q-X.L and C.H.Y.

\section*{Acknowledgements}
H.L acknowledges financial support from the National Natural Science Foundation of China (Grant
Nos. 61803266,61703281,91846301,71790615), Guangdong Province Natural Science
Foundation (Grant Nos. 2019A1515011173,2019A1515011064,2017B030314073), Shenzhen Fundamental Research-general project ( JCYJ20190808162601658). C.H.Y acknowledges the Research Grants Council of the Hong Kong Special Administrative Region, China (Projects No. EdUHK ECS 28300215, GRF 18304316, GRF 18301217 and GRF 18301119), the EdUHK FLASS Dean's Research Fund IRS12 2019 04418, ROP14 2019 04396, and EdUHK RDO Internal Research Grang RG67 2018-2019R R4015.

\section*{References}

\bibliography{bibliography}

\end{document}